\documentclass[preprint]{aastex}
\usepackage{lscape}
\usepackage{emulateapj5}

\begin{document}
\submitted{To appear in the July 2002 issue of The Astronomical Journal}


\title{Searching for Planets in the Hyades II: Some Implications of \\ Stellar 
    Magnetic Activity \altaffilmark{1}}

\altaffiltext{1}{Data presented herein were obtained at the 
W.M. Keck Observatory, which is operated as a scientific partnership among
the California Institute of Technology, the University of California and 
the National Aeronautics and Space Administration. The Observatory was made
possible by the generous financial support of the W.M. Keck Foundation.}


\author{Diane B. Paulson}
\affil{Astronomy Department, University of Texas, Austin, TX 78712}
\email{apodis@astro.as.utexas.edu}

\author{Steven H. Saar}
\affil{Center for Astrophysics, 60 Garden Street, Cambridge, MA 02138}
\email{ssaar@cfa.harvard.edu}

\author{William D. Cochran}
\affil{McDonald Observatory, University of Texas, Austin, TX 78712}
\email{wdc@astro.as.utexas.edu}

\and

\author{Artie P. Hatzes}
\affil{Th\"{u}ringer Landessternwarte Tautenburg, D-07778 Tautenburg, 
Germany}


%\altaffiltext{1}{Visiting Astronomer, Cerro Tololo Inter-American Observatory.
%CTIO is operated by AURA, Inc.\ under contract to the National Science


\begin{abstract}
The Hyades constitute a homogeneous sample of stars 
ideal  for investigating the dependence of planet formation on the mass of 
the central star. Due to their youth, 
Hyades members are much more chromospherically active than 
stars traditionally surveyed for planets using high precision radial velocity 
techniques. Therefore, we have conducted a detailed investigation of 
whether magnetic activity of our Hyades target stars 
will interfere with our ability to make precise radial velocity ($v_{\rm rad}$)
searches for substellar companions.
We measure chromospheric activity (which we take 
as a proxy for magnetic activity) by computing the equivalent of the 
$R^{\prime}_{\rm HK}$ activity index (which 
is corrected for photospheric contributions), from 
the Ca {\sc ii} K line.  $\langle R^{\prime}_{\rm HK}\rangle$ 
is not constant in the Hyades:
we confirm that it decreases with increasing temperature in the F stars,
and also find it decreases for stars cooler than mid-K.
We examine correlations between simultaneously measured 
$R^{\prime}_{\rm HK}$ and radial velocities 
using both a classical statistical test
and a Bayesian odds ratio test. We find that there is a significant 
correlation between $R^{\prime}_{\rm HK}$ and the radial velocity in 
only 5 of the 82 stars in this sample.
Thus, simple $R^{\prime}_{\rm HK}$ -- $v_{\rm rad}$
correlations will generally not be effective in correcting the measured
$v_{\rm rad}$ values for the effects of magnetic activity in the Hyades.
 We argue that this implies 
{\itshape long timescale} activity variations (of order a few years; i.e., 
magnetic cycles or growth and decay of plage regions) 
will not significantly hinder our search for planets in the Hyades if
the stars are closely monitored for chromospheric activity.  
The trends in the radial velocity scatter ($\sigma'_v$) 
with $\langle R^{\prime}_{\rm HK}\rangle$, $v \sin i$, and $P_{\rm rot}$ 
for our stars is 
generally consistent with those found in field stars in the Lick planet 
search data, with the notable exception of a shallower dependence 
of $\sigma'_v$ on $\langle R^{\prime}_{\rm HK}\rangle$ for  F stars.

\end{abstract}

\section{Introduction}

%for decades \citep{hen61,lyn68,spi85}, but
%central brightness peak in M15 \citep{kin75,new78}
%seemed a unique anomaly.  Then \citet{aur82} suggested a central peak
%in NGC 6397, and a limited photographic survey of ours \citep[Paper I]{djo84}

The high-precision radial velocity ($v_{\rm rad}$) technique (e.g., 
\citet{BuMaWi96}) has been remarkably successful in discovering 
planetary-mass companions to nearby main-sequence stars. Studying these systems 
enables us to begin to place important new constraints on the planet formation 
process. 
So far, high precision radial velocity surveys have primarily targeted 
the brightest, and hence the nearest chromospherically inactive, 
slowly-rotating solar-type stars.
Thus, while the targets cover a wide range of ages, compositions, 
and perhaps histories, they are dominated by 
old, higher mass stars.

A more complete understanding of the planet formation process hinges upon 
determining which stellar properties influence planetary system formation and 
dynamics. This can be accomplished with a well controlled and homogeneous sample of target stars. 
By varying only one stellar parameter at a time, we may begin to understand 
the properties and processes which govern planet formation.

We have been searching for planets in the Hyades cluster since 1996 
(\citet{CoHaPa02}, hereafter Paper I). The Hyades cluster 
provides a uniform sample of stars in age and initial metallicity. Stellar 
mass is the main independent variable among the members. 
Thus, if 
planetary systems are found, we will easily be able to study the role stellar mass 
plays in the planet formation process.

In order to detect planetary systems, the $v_{\rm rad}$ signature must be 
unambiguous, and we must be certain it 
is due to an orbiting body and not intrinsic to the star. Several groups have 
explored the implications of intrinsic stellar properties, in particular stellar
pulsations \citep{GrHa97}, rotational velocity, and chromospheric 
activity \citep{SaDo97,SaBuMa98,SaMaNa00a} on radial velocity measurements. 
Stellar non-radial (as well as radial) pulsation can easily give 
observable radial velocity variation, but may be distinguished from orbital 
motion by their resulting variations in stellar line bisectors \citep{GrHa97,
HaCoJo98,HaCoBa98a,HaCoBa98b,BrKoHo98a,BrKoHo98b}.
The rotational modulation and evolution 
of velocity structures or non-uniform active regions 
can cause similar $v_{\rm rad}$ signatures to orbiting planets. 
\citet{WaBoWa92} showed that in 
the case of $\gamma$ Cep, the $v_{\rm rad}$ measurements suggestive 
of a planet were instead likely due to rotation of 
active regions seen in equivalent widths of the Ca {\sc ii} 8662 {\AA} line.
\citet{WaWaIr95} detected a correlation between $v_{\rm rad}$ and long-term Ca 
{\sc ii} emission variations in $\kappa^1$ Ceti probably due to its 
magnetic cycle.  \citet{QuHeSi01} 
showed that the apparent radial velocity signal in HD 166435, which was
originally interpreted as due to a planetary companion 
is actually correlated with photospheric line profile variations and is 
best explained by the presence of dark photospheric spots. 
\citet{DrNo90} discussed the possibility of apparent false 
Doppler shifts over the course of a stellar activity cycle, though, 
\citet{McMoPe93} has shown that solar cycle magnetic variations are undetectable
in disk integrated spectra of the sun.
Saar and collaborators 
\citep{SaHaCo02,PeSaHa00,SaFi00,SaSn99} are working on methods 
for correcting stellar radial velocities to compensate for the stellar 
activity, but these techniques are still under development.  
Because of worries about such activity-related $v_{\rm rad}$ noise,
past planet surveys have been biased against young, active stars 
\citep{Vo02,HeBaDo97, SaDo97}. Consequently, out of 76 planets
discovered as of December 2001, only four have been found 
around young, active stars: 
$\epsilon$ Eridani \citep{HaCoMA00}, $\iota$ Horologii 
\citep{KuEnEl00}, HD 192263 \citep{SaMaNa00a}, and
GJ 3021 \citep{NaMaPe01}. 

Because the Hyades cluster is so young (625 $\pm$ 50 Myr; \citet{PeBrLe98}),
the stars can be quite active. Thus, we must determine if 
this activity is a major contributor to the observed $v_{\rm rad}$ variances. 
In this paper, we explore the connection between activity variations 
and radial velocity measurements in Hyades stars by looking 
for correlations between the Ca {\sc ii} K emission and the $v_{\rm rad}$
(measured in the same spectra).
%activity will correlate with those in the $v_{\rm rad}$. 
This will help us 
understand to what degree this phenomenon masks or complicates detection 
of the low velocity signal induced by planetary mass companions.

\section{Observations and Analysis}
We are conducting a radial velocity survey of Hyades dwarfs using the Keck 
HIRES spectrograph to search for Jupiter and Saturn-mass companions (see Paper I).
By using a 10m class telescope, we are able to 
study even low mass Hyads (down to $M \approx 0.4M_\odot$).
So far, this survey has produced a number 
of stars showing rms velocity variation significantly larger than the 
internal velocity uncertainty ($\sigma_i$,  which includes instrumental noise, 
photon noise, and increased noise due to observational effects such as reduced 
line density and depth in F stars and increased line width in stars with 
higher $v \sin i$). 
Unfortunately, the scheduling of our observing time on Keck has made 
it impossible to sample short periods without significant aliasing. We 
began observations of short period candidates with the High Resolution 
Spectrograph on the Hobby-Eberly Telescope in the fall of 2001.

\subsection{Sample}
Our total sample consists of 98 Hyades dwarfs ranging from F5 to M2. All of the 
targets are confirmed members according to \citet{PeBrLe98}, and 
all known spectroscopic binaries have been removed. We imposed a 
stellar rotational velocity limit of $v \sin i$ $\le$  15 km s$^{-1}$. For the 
chromospheric activity study presented here, we neglect all stars with fewer 
than six observations, reducing the sample to 82 stars.

\subsection{$v_{\rm rad}$ measurements}
The observations at Keck I make use of the HIRES \citep{VoAlBi94} with
I$_{2}$ gas absorption cell as a standard velocity reference \citep{VaBuMa95} 
as detailed in Paper I. 
S/N$\sim$300 is required in this configuration (at 
resolution $R \simeq$ 60,000) to obtain $v_{\rm rad}$ 
precision (in the best cases) of $\sim$3 m s$^{-1}$ (see Paper I); S/N$\sim$100 yields precision of 
$\sim$5-6 m s$^{-1}$. In 
addition, the exposure times must be kept to $\leq$15 minutes. 
We use standard IRAF packages to reduce the CCD images and extract the observed 
spectra. 

It is useful to explore how the rotational broadening in the spectra affects
the internal $v_{\rm rad}$ errors ($\sigma_i$).
Since we have preselected stars
with $v \sin i$ $\le$ 15 km s$^{-1}$, we only have a small range of $v \sin i$ 
to study. In
Figure 1, we show the relationship between $v \sin i$ (taken from the literature
as referenced in Table 1) and the mean internal errors ($\langle\sigma_i \rangle$)
we achieve. Stars with only $v \sin i$ upper limits have not been included in 
the analysis. Error bars on $\langle \sigma_i \rangle$ indicate the rms about
the mean. 
We note that even with this limited set of data, we see the trend that 
one would expect; that increasing $v \sin i$ degrades our velocity precision,
due to the lowered precision in determining the center of broader lines.
In order to estimate how large our errors are at $v \sin i$=15 km s$^{-1}$, 
we fit a linear relation to the data. We find 

\begin{equation}
\langle \sigma_i \rangle=1.050 + 0.555*\it v \sin i \ \ \rm m \ \rm s^{-1}.
\end{equation}
Therefore, at the cutoff of 15 km s$^{-1}$, we can still 
achieve $v_{\rm rad}$
precision of $\sim$ 10 m s$^{-1}$. We 
would like to point out that these are our {\itshape mean} internal errors of all of our 
spectra of a given star. If instead we had chosen 
the highest S/N spectrum, we would have derived $\sigma_i$ = 6.5 m s$^{-1}$
at $v \sin i$ = 15 km s$^{-1}$. For $v \sin i \la$ 3 km s$^{-1}$, other sources
of line broadening become important (e.g., macroturbulence), and 
$\langle \sigma_i \rangle$ is no longer dominated by $v \sin i$. 

\subsection{Calibration of Chromospheric Activity}

Most stellar chromospheric activity is attributed to the interaction of 
magnetic fields with 
convection \citep{MiZw81,TiZw81,Mi82}. However, the effect of magnetic 
processes on the integrated radial velocity of a star at the
few m s$^{-1}$ level is only beginning to be explored. To date, stars which 
are chromospherically active and show no obvious trend in the velocity signal
over a given timescale are either observed less frequently
or removed from planet 
surveys \citep{Vo02,CuMaBu99, SaDo97}. This is because these stars tend to show 
somewhat higher levels of radial velocity scatter, attributed to the magnetic 
activity, and this will possibly inhibit detection of low-amplitude velocity 
signals.

The question arises as to whether stellar activity in 
Hyades  stars induce significant periodic 
centroid shifts in photospheric absorption lines,
and thus $v_{\rm rad}$ signals which could be confused for
perturbations made by planetary companions.
The members 
of the Hyades cluster are chromospherically active, and quite 
young. 
Because activity might be a significant source of radial velocity variation 
\citep{SaDo97, SaBuMa98, SaFi00}, we 
include the Ca {\sc ii} H \& K lines in the spectral region of each exposure 
for velocity measurement with the Keck HIRES. We have monitored the 
chromospheric activity of our target stars, acquiring Ca {\sc ii} K 
emission core flux measurements simultaneously with each velocity measurement.

To measure stellar chromospheric activity, the Mt. Wilson $S$ index is adopted. 
This index is defined  
(e.g. \citet{BaDoSo95}) as a quantity proportional to the sum of 
the flux in 1{\AA} FWHM triangular 
bandpasses centered on the Ca {\sc ii} H and K 
lines divided by the sum of the flux in 20{\AA} bandpasses in the continuum at 
3901 and 4001{\AA} \citep{SoDuJo91, BaDoSo95}.

At Keck 1, we set HIRES so that the Ca {\sc ii} H and K lines are 
contained within 
the spectra for all stars observed. The four quantities to be measured (the 
two calcium line core fluxes plus 
the two continuum bandpass fluxes) are spread across three echelle spectral orders 
which overlap by a few {\AA}. As a result of flat fielding uncertainties 
in the most blueward order, 
we do not use measurements in this order, i.e. of the blue (3901{\AA}) 
continuum.  
In addition, we do not measure the Ca {\sc ii} H line (at 3968.47{\AA}) because 
for some stars, the wings 
of strong Balmer H$\epsilon$ features (at 3970.07{\AA}) lay within the measured 
bandpass. This would affect measurement of the Ca {\sc ii} H line flux as the 
absorption can be strong enough so as to artificially decrease the Ca {\sc ii} H
 flux,
and strong H$\epsilon$ emission (seen in a few of the M dwarfs) will
artificially increase 
the measured Ca {\sc ii} H line flux. Therefore, we have defined an index 
$S_{\rm Keck}$ which is the 
ratio of the flux in a 1{\AA} triangular bandpass centered on the Ca {\sc ii} K 
line to the 
flux in a 20{\AA} bandpass centered on the redward continuum at 4001~{\AA}. 
Figure 2 shows the numerical filter we used in measuring the Ca {\sc ii} K flux.
 Several 
of our Hyades program stars have previously been measured as part of the Mt. 
Wilson program \citep{DuFrLa84}. In Figure 3, we show the 
relationship between our $S_{\rm Keck}$ and $S_{\rm Duncan}$ 
(the published mean $S$ data 
for stars measured by \citet{DuFrLa84} including error bars
which characterize the rms of the measured $S$ index over the course of 
observations (including variability). 
We find the linear relation 
\begin{equation}
S_{\rm Duncan}=-0.027(\pm 0.046) + 0.991 (\pm 0.120) S_{\rm Keck}
\end{equation}
\noindent 
Both the slope and the intercept of the formal fit are within 1 $\sigma$ of 
$S_{\rm Keck}=S_{\rm Duncan}$. We are thus able to transform our data 
into a standard Mt. Wilson $S$ index scale using 
Equation 2. We call our transformed values $S_{\rm MW}$.

In addition, we further transform our chromospheric $S$ index to 
$R^{\prime}_{\rm HK} = F^{\prime}_{\rm HK}/F_{\rm bol}$, where 
$F^{\prime}_{\rm HK}$ is the Ca {\sc ii} HK surface flux, corrected for
flux contributions from the photosphere, and $F_{\rm bol}$ is the bolometric
flux \citep{NoHaBa84,Mi82}. Thus, $R^{\prime}_{\rm HK}$ is the normalized, 
purely chromospheric component of the HK flux and is also useful as 
a means to compare stars of different spectral types. 
To convert from $S_{\rm MW}$ to $R^{\prime}_{\rm HK}$, 
we used the method outlined in \citet{NoHaBa84}, with $B-V$ 
color index of each star obtained from \citet{APLa99}, 
and no reddening corrections were used. 

\section{Results}

\subsection{Chromospheric activity}
Figure 4 shows the variation of $\langle R^{\prime}_{\rm HK}\rangle$ 
with $B-V$ for our sample of Hyades dwarfs. 
This demonstrates that activity as measured by 
$\langle R^{\prime}_{\rm HK}\rangle$ is
{\it not} constant with $B-V$ in Hyades stars:  $\langle R^{\prime}_{\rm HK}\rangle$ 
values decrease 
from a plateau with $\langle \log R^{\prime}_{\rm HK}\rangle \approx -4.42$ 
for $B-V<$0.7 and for $B-V>$1.1.
\citet{NoHaBa84} notes that the values of 
$R^{\prime}_{\rm HK}$ for $B-V>$1.0 may be uncertain by as much as 20\%.
However, the downward trend of $B-V$ below 0.7 has not yet been addressed to our
knowledge. A few outliers with $B-V<$0.7 (see Fig. 4) were determined not 
to be Hyades members 
by D. Latham (private communication) based on too large Hipparcos distances, 
photometry below main sequence and velocities which disagree with members, 
in contrast to \citet{PeBrLe98} who found these same stars to be members.
The decline for $B-V<$0.7 confirms the trend observed by \citet{DuFrLa84} 
and \citet{So85} and seen in the data collected by \citet{Ru87}.  HK data for 
the younger Pleiades cluster (age $\sim 0.1$ Gyr)
shows more scatter \citep{Ru87}, but a similar, though 
weaker, decline for $B-V<$0.7.
This suggests that the decline in activity for $B-V<$0.7 
is real (as opposed to a problem in $R^{\prime}_{\rm HK}(B-V)$ calibration);
a color-dependent calibration error should not evolve significantly
between the ages of the Pleiades and Hyades.
Thus, although we must be cautious about stars with $B-V>$1.0,  
we confirm that at fixed age, there is a broad maximum in chromospheric 
emission for 0.7$\leq B-V \leq 1.0$ ($\approx$G5 to K3), with decreases
for both hotter and (tentatively) for cooler stars.  

\subsection{Intrinsic $v_{\rm rad}$ Jitter}

\subsubsection{Comparison of Observed $v_{\rm rad}$ Jitter with Empirical Relations}

\citet{SaBuMa98} (hereafter SBM) used the observed radial velocity dispersion 
($\sigma_{v}$) in the Lick planet search data to compute the $v_{\rm rad}$ 
dispersion above the internal errors,
$\sigma^{\prime}_{v} = (\sigma_{v}^2-\sigma_i^2)^{0.5}$.  They then searched 
for correlations between $\sigma^{\prime}_{v}$ and 
$\langle R^{\prime}_{\rm HK}\rangle$, rotational period
($P_{\rm rot}$), and projected rotational velocity ($v \sin i$).
They also developed a simple model based on rotating/evolving starspots
and plage to predict $\sigma^{\prime}_{v}$.
The Hyades dwarfs studied here are significantly more active on average than
those stars studied in SBM, making it useful to investigate
whether Hyad $\sigma^{\prime}_{v}$ behave similarly.
We calculated $\sigma^{\prime}_{v}$ in the same manner as
SBM, removing internal noise from each measurement.
Due to the selection of stars with $v \sin i \leq$15 km s$^{-1}$ (Paper I),
we have also selected for stars with low radial velocity jitter as shown in the
model developed in SBM. We have also ignored stars with very large 
($\gg$ 100 m s$^{-1}$) systematic velocity trends. F stars in our sample (ignoring one outlier) 
have $\sigma^{\prime}_{v}$ ranging from $\sim$8 to 38 m s$^{-1}$ 
with an average value of $\langle \sigma^{\prime}_{v}\rangle = 14.8\pm8.4$ m s$^{-1}$,
G stars range 
from $\sim$5 to 40 m s$^{-1}$ ($\langle \sigma^{\prime}_{v}\rangle = 18.9\pm9.0$ 
m s$^{-1}$), K stars (ignoring one outlier) range 
from $\sim$3 to 45 m s$^{-1}$ ($\langle \sigma^{\prime}_{v}\rangle = 14.2\pm13.6$
m s$^{-1}$ , and M stars from $\sim$4 to 20 m s$^{-1}$ 
($\langle \sigma^{\prime}_{v}\rangle = 12.7\pm7.6$ m s$^{-1}$).

A comparison of the observed $\sigma^{\prime}_{v}$ and the 
values predicted by empirical relationships found in the Lick
$v_{\rm rad}$ database 
($\sigma^{\prime \rm (pred)}_{v}$; see SBM) are shown in  Fig. 5 (first three
panels) and summarized in Table 2.  For the rotation-related  quantities,
$P_{\rm rot}$ and $v \sin i$, agreement is generally quite good,
with the scatter $\sigma$ about the line 
$\sigma'_v = \sigma^{\prime \rm (pred)}_{v}$ at or
below the values found for the empirical SBM fits.
(We have considered only stars with measured $P_{\rm rot}$.)
Stars with only upper limits to $v \sin i$ are also all in agreement
with predictions.

Only relatively few Hyades stars have measurements of
$P_{\rm rot}$ and $v \sin i$ needed
to compare $\sigma'_v$ with the predictions of $\sigma^{\prime \rm (pred)}_{v}$
based on these parameters using the SBM results.
With $\langle R'_{\rm HK}\rangle$ though,
all the Hyades sample can be used except M stars
(which were not studied in SBM).
If $\sigma^{\prime \rm (pred)}_{v}$ is predicted
using $R'_{\rm HK}$ and the combined Lick G and K star sample,
the scatter $\sigma_{rms}$ of Hyades G and K stars' $\sigma'_v$ about this
$\sigma^{\prime \rm (pred)}_{v}(R'_{\rm HK}$(G+K))
is slightly higher than for the Lick stars. This is true
even if one Hyades outlier (a K star) is removed ($\sigma_{rms}$(Hyades) =
0.28 dex compared with $\sigma_{rms}$(Lick) = 0.25).
To explore this result further, we compared $\sigma'_v$ for
the Hyad G and K stars separately
with $\sigma^{\prime \rm (pred)}_{v}(R'_{\rm HK}$(G+K)),
and also with $\sigma^{\prime \rm (pred)}_{v}$
based on separate empirical fits to the Lick G and K stars
(not published in SBM).
These tests indicate that $\sigma_{rms}$(Hyades) is smaller than
$\sigma_{rms}$(Lick)
for the G stars ($\approx$0.22 dex), whether $\sigma'_v$ is compared to
$\sigma^{\prime \rm (pred)}_{v}(R'_{\rm HK}$(G+K)) or
$\sigma^{\prime \rm (pred)}_{v}(R'_{\rm HK}$(G)).  The scatter
is notably worse for K stars, though improved somewhat with
the separate K star empirical fit:
$\sigma_{rms} \approx$0.32 dex using
$\sigma^{\prime \rm (pred)}_{v}(R'_{\rm HK}$(G+K)),
$\sigma_{rms} \approx$0.28 dex using
$\sigma^{\prime \rm (pred)}_{v}(R'_{\rm HK}$(K)), with one outlier
removed in each case.  Thus, since F, G, and K stars are
best fit when treated separately, the Hyades
$\sigma^{\prime}_{v}$ data suggest a systematic spectral type dependence
in the relationship between $\sigma^{\prime}_{v}$ and
$R'_{\rm HK}$.

The Hyades F stars (excluding one outlier) show a systematic
trend relative to the $\sigma'_v = \sigma^{\prime \rm (pred)}_{v}$ line
which suggests the
empirical relation ($\sigma'_v$(F)$\propto \langle R'_{\rm HK}\rangle^{1.7}$;
SBM) may be too steep.  Indeed, the scatter is lower
and the trend largely removed if the relations for G or G+K stars are used
instead ($\sigma_{rms} = 0.21$ dex or 0.28 dex, respectively,
compared with $\sigma_{rms} = 0.38$ dex using the F star
fit).  The $v \sin i \leq 15$ km s$^{-1}$ limit of our sample, though,
may be a factor; while
this limit has little effect on G and K Hyades stars, it definitely
excludes some high $v \sin i$ (and hence high $R'_{\rm HK}$)
F stars. The Lick analysis was performed without any $v \sin i$
restrictions, and includes two F stars with $v \sin i \geq 15$ km s$^{-1}$.
Restrictions on $v \sin i$ will tend to limit $\sigma'_v$ for a fixed
$R'_{\rm HK}$, since the higher $v \sin i$ is an important component
for enhanced $v_{\rm rad}$ noise \citep{SaDo97}.
 
%On the other hand, some of the Lick F stars may have had undiscovered
% planets, skewing the $\sigma'_v$ -- $R'_{\rm HK}$ relationship with their
%planet-enhanced $\sigma'_v$.

The two ``outlier'' stars noted above (one F and one K type) show
$\sigma'_v$ values significantly enhanced
(by $>2\sigma_{rms}$(Lick)) over that predicted by
$\sigma^{\prime \rm (pred)}_{v}(R'_{\rm HK})$.
Hence, these stars have
large additional $v_{\rm rad}$ variations not associated with activity,
and are particularly good candidates for further
and more frequent observations in search of low mass companions.

\subsubsection{Comparison of Observed $v_{\rm rad}$ Jitter with a Simple Model}

We have also compared Hyades $\sigma^{\prime}_{v}$ values with the combined
spot/plage $\sigma'_v$ model  outlined in SBM.  This model
combines the rotating spot model from \citet{SaDo97} with a (very) simple
plage model which assumes that convective velocity changes due to plage
magnetic fields are $\propto v_{\rm mac}$,
the macroturbulent velocity. The full model is given by
{\scriptsize
\begin{equation}
\sigma^{\prime {\rm (mod)}}_v \approx
   \sqrt{[4.6 f_S^{0.9} (v \sin i)\cos \langle \theta \rangle ]^2
  + [\alpha A(f_P) (v_{\rm mac} + v \sin i)]^2} {\rm [m s}^{-1}]. 
\end{equation}
}
\noindent 
Here, $f_S$ is the {\it differential} spot filling factor (the portion
of the total spot filling factor responsible for
photometric variations). Also, 
$\langle \theta \rangle$ is the mean spot latitude, 
$\alpha$ is an adjustable  constant, and $A(f_P)$ is a function of the
plage filling factor $f_P$, given by
$A(f_P) = 0.25 \sin (\pi\sqrt{f_P})$ (the scaling factor was not 
explicitly given in SBM; it scales the maximum of 
$A(f_P) = f_P$).  Following SBM, we take 
$f_P = 0.08 (\tau_c/P_{\rm rot})^{1.8} < 0.65$ 
(\citet{Sa96}; where $\tau_c$ is the
convective turnover time from \citep{NoHaBa84}), $\langle \theta \rangle 
= 45^\circ$, $\alpha \approx 9$, and 
estimate $v_{\rm mac}$ from relations for active stars in \citet{SaOs97}.

Unfortunately, only five stars had the required data ($v \sin i$, $P_{\rm rot}$,
 and photometry sufficient to estimate $f_S$) to make a 
complete model estimate of $\sigma'_v$.  
If we assume $f_S \approx 0$ for F stars (Hyades F stars show
little photometric variability -- \citet{RaLoSk98}) used them as 
constant check stars), and estimate their $P_{\rm rot}$
from $R'_{\rm HK}$ (vis. \citet{NoHaBa84}) we gain two more stars. 
Several more have upper limits to $v \sin i$. As shown in
Figure 5 (lower right) and Table 2,  the agreement between $\sigma'_v$ 
and $\sigma^{\prime \rm (mod)}_v$ for the
(few!) stars we can model adequately is quite good: $\sigma_{rms}$ is
$\approx$6 to 7 m s$^{-1}$ or $\approx 0.13$ dex (comparable to 
SBM). All cases where only upper limits on
$\sigma'_v$(mod) are possible (due to $v \sin i$ upper limits) 
are in agreement as well.

\subsection{$R^{\prime}_{\rm HK}$ vs. $v_{\rm rad}$}

We have calculated the linear correlation coefficient ($r$) in the standard way 
\citep{BeRo92} to search for any statistical 
correlations between the chromospheric activity and the radial velocity for each star. The 
values of $r$ (which range from -1 to 1) calculated are shown in column 6 of Table 3. 
We then use this quantity 
to test statistically (both in a frequentist and a Bayesian manner) whether any
linear correlations exist between the $v_{\rm rad}$ and $R^{\prime}_{\rm HK}$.

We first follow the frequentist approach given in \citet{BeRo92} 
to determine a probability that the $v_{\rm rad}$ and $R^{\prime}_{HK}$ are 
correlated. The 
correlation coefficient is used to find a probability $P_{\rm c}$
that the data come from an uncorrelated parent population 
($P_{\rm c}$=1 indicates that the data are completely uncorrelated, while 
$P_{\rm c}$=0 indicates be a completely correlated data set). 
The results from this test are in column 8 of Table 3. 
In using this method, we find several stars with high 
probabilities that the data come from a correlated parent sample. However, 
this method introduces a bias, since 
$P_{\rm c}$ is calculated by only considering $r$  relative to 
an uncorrelated parent sample. 
What is perhaps a better test is to compare the data to both
correlated and uncorrelated parent samples.
This way, we can test both hypotheses and make a better determination of 
possible correlation in the parent sample. To do this, we use 
a Bayesian approach.

Our Bayesian analysis uses an odds ratio ($K$) defined by \citet{Je61}.
This incorporates prior knowledge of a probability distribution of 
$r$ of the parent sample. Here, we choose a prior 
distribution to be constant and centered on (0,0).  We calculate a ratio of the 
probability that $v_{\rm rad}$ and $R^{\prime}_{HK}$ are uncorrelated versus the 
probability that they are correlated (i.e., $K\ll$1 indicates a  
strong correlation, $K\gg$1 the lack of one).
For example, for $K=2$, there are 2:1 odds that the sample is 
uncorrelated (i.e., a 67\% chance that the sample is uncorrelated). 
Column 9 of Table 3 gives the odds ratios. However, \citet{Sc69} notes that the strength of correlations worked 
out in this manner is somewhat overestimated: 
correlation probabilities $<$75\% 
are highly suspicious. In contrast to the 
frequentist correlation analysis, with the Bayesian 
techniques, we show that 15 stars have greater than 50\% chance of being 
correlated.
Of these, 5 show a strong chance of correlation- greater than 70\%. 
Due to the small number of observations, Bayesian statistics will yield more reliable results, so these 
values are the ones the authors favor. Of the remaining stars, 40 show slight
non-correlation between $v_{\rm rad}$ and $R^{\prime}_{HK}$ and 11 show slight 
correlation. So, the overwhelming majority of stars are slightly to strongly
uncorrelated.
Figures 6 and 7 show examples of stars with uncorrelated and correlated trends,
respectively, from the previous analysis.

\section{Discussion}

Our fundamental result is that very few Hyads (5 of 82)
show significant correlations between simultaneous $v_{\rm rad}$ and $R'_{\rm HK}$ 
measurements.  It is therefore important 
to review how stellar activity can alter
observed radial velocities, and what our result then implies
for planet detection in the Hyades.   In the Sun, Ca {\sc ii} emission
primarily reflects the surface coverage of plage and active network,
since it is relatively weak over sunspots themselves (e.g. \citet{LiAv70} and 
references therein), and spot area is typically small relative to plage area. 
Thus, $v_{\rm rad}$ fluctuations due to spots (e.g., \citet{SaDo97}) will not have
corresponding $\Delta R'_{\rm HK}$, and so spots should not be a
significant contributor to $v_{\rm rad}$ -- $R'_{\rm HK}$ correlations.
Plage/network can generate $v_{\rm rad}$ fluctuations 
in two main ways. First, plage is slightly brighter in the continuum 
(by a few \%) than the quiet Sun.  Assuming this is also 
true for stars, rotation of
inhomogeneous patches of plage will cause traveling enhancements at the local 
intensity-weighted rotational velocity, which will
translate into apparent $v_{\rm rad}$ changes on $P_{\rm rot}$ timescales. 
This effect is completely analogous to the one caused by spots. 
Since the 
brightness enhancement of plage is tiny compared with the $\sim$90\% 
light deficit (in $V$) due to sunspots, though, the brightness
effect of plage on $v_{\rm rad}$ should be small, 
even considering the typically larger plage area. 

To explore the effects on $v_{\rm rad}$-activity correlations, consider that 
an identical plage, observed at two rotation angles
$\pm \phi$ (measured from disk center), will exhibit identical 
activity enhancements (i.e.,  $\Delta R'_{\rm HK}(\phi) = 
\Delta R'_{\rm HK}(-\phi)$), 
but will show perturbations to $v_{\rm rad}$ due to brightness
of opposite sign: $\Delta v_{\rm rad}(\phi) = - \Delta v_{\rm rad}(-\phi)  = \alpha 
v_{\rm rot}(\phi)$
(where $\alpha$ is some function of the plage brightness enhancement).  
Thus, a given plage will show a scatter of $\Delta v_{\rm rad}$ due to its
brightness as it rotates across the disk.
% *** HERE  (variable activity?)
When averaged over enough observations, intensity changes due to 
activity (spots or plage) have no net $v_{\rm rad}$ effect, since the 
perturbations $\propto \pm v_{\rm rot}(\phi)$ will average to zero
over the many observed $\phi$. Changing the mean level of
activity will change the average magnitude of the effect and its rms, 
but once again, for sufficient $\phi$ coverage the net
$\Delta$$v_{\rm rad}$ = 0.

In addition to its brightness perturbation, plage also induces 
changes in the local velocity field, 
suppressing convective velocities in strong magnetic fields.  
The altered velocity field induces changes in the 
line shape \citep{Li82}  
and leads to an overall convective blueshift in the Sun 
\citep{CaCeRi85}. Furthermore, these changes will
vary from line to line based on their strength and excitation, just as 
line bisectors do (e.g., \citet{AsNo00}).  
%
%The $\Delta$$v_{\rm rad}$ due to the plage's altered velocity field are more complex.
Thus, in the case of plage the $v_{\rm rad}$ perturbation may be written
$\Delta$$v_{\rm rad}$$(\phi) =v_{\rm blue}(\phi)+v_{\rm shape}(\phi)$,
where $v_{\rm blue}(\phi)$ is the overall convective shift
of the line core, and $v_{\rm shape}(\phi)$ is the
apparent $v_{\rm rad}$ change due to the altered line profile shape.
The latter arises because most 
methods measure the $v_{\rm rad}$ of individual exposures by comparison to  
a single high S/N ``template'' spectrum (either an average spectrum
or a single deep exposure).  Fluctuations in 
line shape relative to this ``template'' will inevitably lead to an
apparent shift in the line centroid, and hence $v_{\rm rad}$.
%as well as from the plage's position ($\propto v_{\rm rot}(\phi)$).  
%Once again, on short ($\leq P_{\rm rot}$) timescales, 
%the underlying $v_{\rm rot}(\phi)$  at the plage's position
%will cause a $\pm$ fluctuation to the plage induced offset (which itself
%may vary in a complex way with $\phi$). This in turn will lead 
%to a scatter of $\Delta$$v_{\rm rad}$ for a given activity change $\Delta R'_{\rm HK}$,
%depending on at what disk position(s) the plage is observed. 
%Over long timescales, this $\pm v_{\rm rot}(\phi)$ will average out, but 
In contrast with brightness perturbations, the time average of 
$v_{\rm blue}(\phi) + v_{\rm shape}(\phi)$ will in general be non-zero, since
both, due to their intimate connection with 
convection are  (unlike $v_{\rm rot}$) symmetric about $\phi=0$
in the time-averaged sense. 
Clearly, {\it strong correlations between $v_{\rm rad}$ and $R'_{\rm HK}$ will
result primarily from these long timescale changes in average plage area}
(see also \citet{SaFi00}).  The present results say little about short term
(timescales $\la P_{\rm rot}$) changes in $v_{\rm rad}$, or about the effects of
starspots.  

The lack of many significant  $v_{\rm rad}$ -- $R'_{\rm HK}$ correlations
in the Hyades thus implies that {\it longterm} changes in plage activity
have little effect on $v_{\rm rad}$ for our stars. This is not an entirely
surprising result.  If we look at active stars in \citet{BaDoSo95}, 
relatively few show clear cycles in Ca {\sc ii} (see also \citet{Sa02}).
Without a systematic long-term activity variation, active stars are 
less likely to show strong $v_{\rm rad}$ -- $R'_{\rm HK}$ correlations. 
Typically, short term (rotational) jitter in $v_{\rm rad}$, and perhaps 
flares in activity, are expected to
dominate active stars without strong cycles,
swamping potential $v_{\rm rad}$ -- $R'_{\rm HK}$ correlations with rapid fluctuations
in one or both variables.   Thus, the method suggested by
\citet{SaFi00} for correcting $v_{\rm rad}$ timeseries for some of the 
jitter induced by activity will not be effective for most Hyads.
Other methods, based for example on variable line bisector changes 
\citep{SaFiSn01,SaHaCo02,QuHeSi01} may be useful in
diagnosing (and possibly correcting) for short term
$v_{\rm rad}$ jitter from spots and plages.

Clearly, the changing $\langle R'_{\rm HK}\rangle$ with $B-V$ implies that the 
inverse Rossby number $Ro^{-1} = \tau_C/P_{\rm rot}$ (at least as defined by 
\citet{NoHaBa84}) is not constant with mass at fixed age. 
This confirms and extends (to $B-V > 1.0$) the similar 
conclusion of \citet{So85}. Consistent with this, the age calibration of
\citet{DoDoBa97}, which estimates age $t$ as a function of
$\langle R'_{\rm HK}\rangle$ alone, predicts 
$\langle \log t$[yr]$\rangle = 8.89 \pm 0.27$ (in agreement 
with \citet{PeBrLe98}) but overestimates $t$ for $B-V < 0.60$ and $B-V > 1.30$. 
The physical implications of a mass dependent
$\langle R'_{\rm HK}\rangle$ in the Hyades are less clear.
Apparently, either generation of magnetic flux,
or the physics of chromospheric heating, or both, vary with mass at fixed
age. There is some independent evidence for both of these  ideas.
\citet{Sa01} finds a monotonic relation between magnetic flux
and $Ro^{-1}$, which implies mass dependent magnetic flux
in the Hyades, given that $Ro^{-1}$ is not constant. Supporting the
concept of mass-dependent changes in heating,
several researchers (e.g., \citet{RuZwSc89}) have noted that in
comparison to hotter stars, M dwarfs show
Balmer emission dominating over HK emission, and chromospheric emission
in general reduced relative to coronal emission.

Taken as a whole, the $\sigma'_v$ values for the Hyades are in reasonably
good agreement with empirical results and models in SBM.  The main  
exceptions are that the Hyades data suggest a more continuous
change in the dependence  of the $\sigma'_v$ - $\langle R'_{\rm HK}\rangle$
relationship  on  spectral type (e.g., G and K stars are better considered 
separately).  Also, the SBM fit for  F stars, $\sigma'_v \propto 
\langle R'_{\rm HK}\rangle^{1.7}$ may be too steep, as also suggested by
\citet{SaMaNa00b}. The difference may however 
be the result of the limitation of $v \sin i \leq 15$ m s$^{-1}$ in the present 
sample.  Further analysis of the Lick and Hyades data, with and without
$v \sin i$ limits, is needed to resolve this issue.

%Like the brightness effect, one result will be $v_{\rm rad}$ fluctuations on 
%$P_{\rm rot}$ timescales.  Unlike the intensity perturbations, 
%though,  {\it long-term} changes in the mean plage area (due to magnetic cycle
%variations, growth/decay of active regions or active longitudes)
%{\it do} have a net $v_{\rm rad}$ effect, since they produce a net 
%change the average line shape and convective blueshift.

%changes in the {\it apparent} disk averaged rotational 
%The paucity of Hyads

\acknowledgments
We wish to thank Debra Fischer for many useful suggestions.
We would also like to thank Chris Sneden and Bill Jefferys for 
several useful discussions on the topics of this paper. In addition, we are 
grateful to David Latham and Robert Stefanik for their assistance in identifying
binary stars and Hyades non-members from our sample.
DBP and WDC are supported by NASA grants NAG5-4384 and 
NAG5-9227 and NSF grant AST-9808980. SHS is supported by NASA 
grant NAG5-10630.




%\bibliography{/home/zorba/rv/apodis/papers/master.bib}

\begin{figure}
\plotone{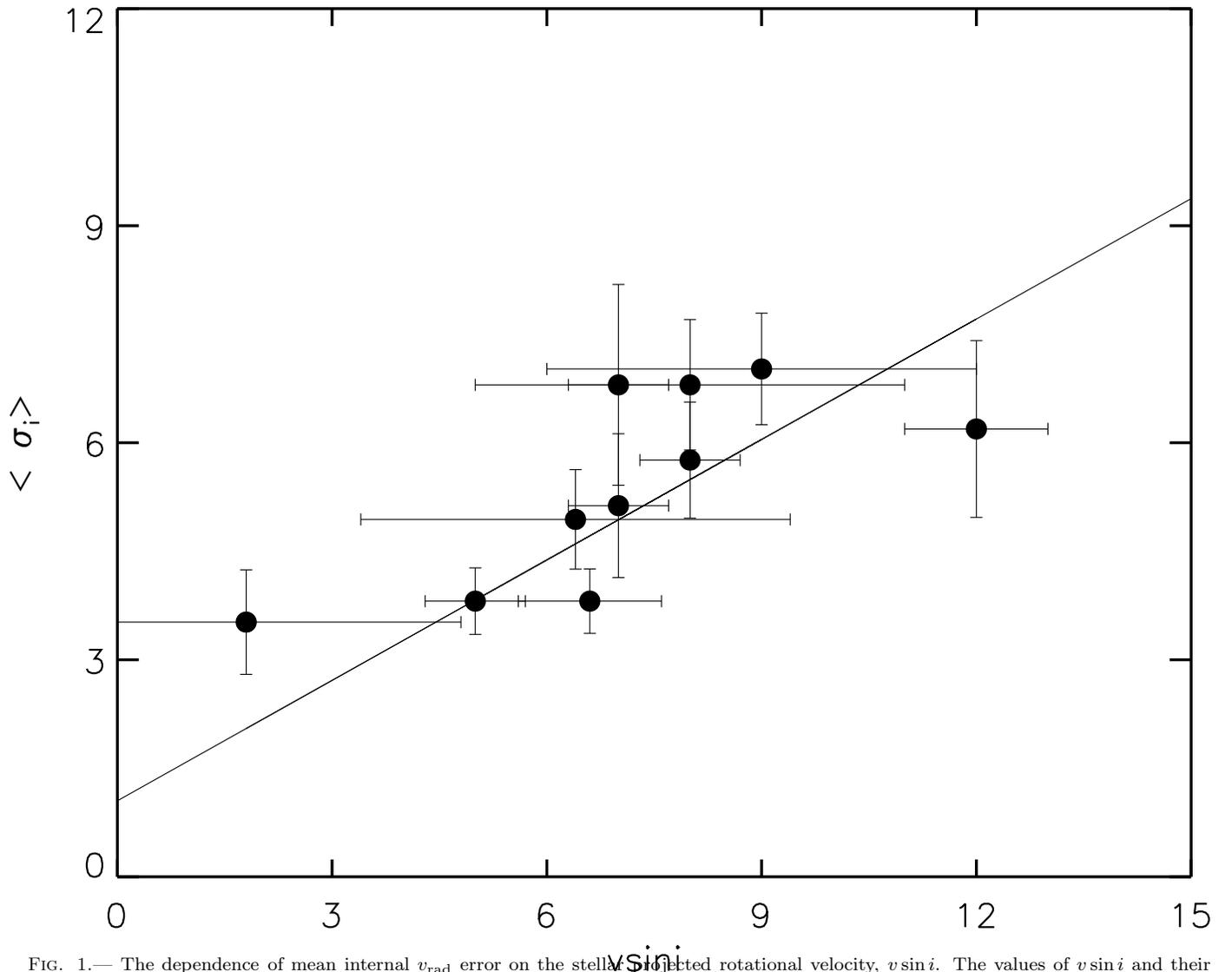}
\caption{The dependence of mean internal $v_{\rm rad}$ error on the stellar projected rotational velocity, $v \sin i$. The values of $v \sin i$ and their 
sources are listed in Table 1. Vertical 
error bars indicate the standard deviation from the mean for all of our velocity measurements for a given star.}
\end{figure}

\begin{figure}
\plotone{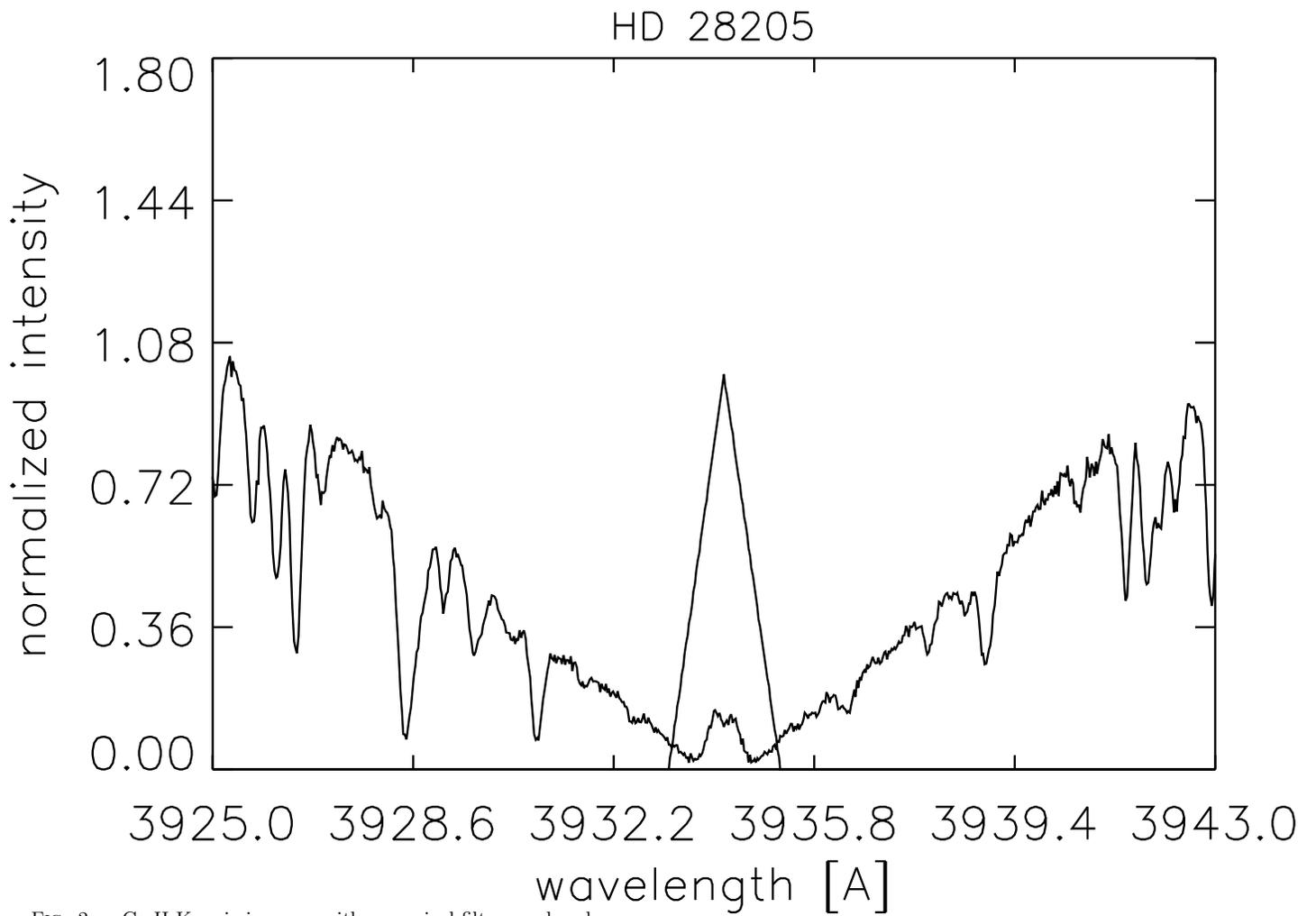}
\caption{Ca II K emission core with numerical filter overlayed}
\end{figure}

\begin{figure}
\plotone{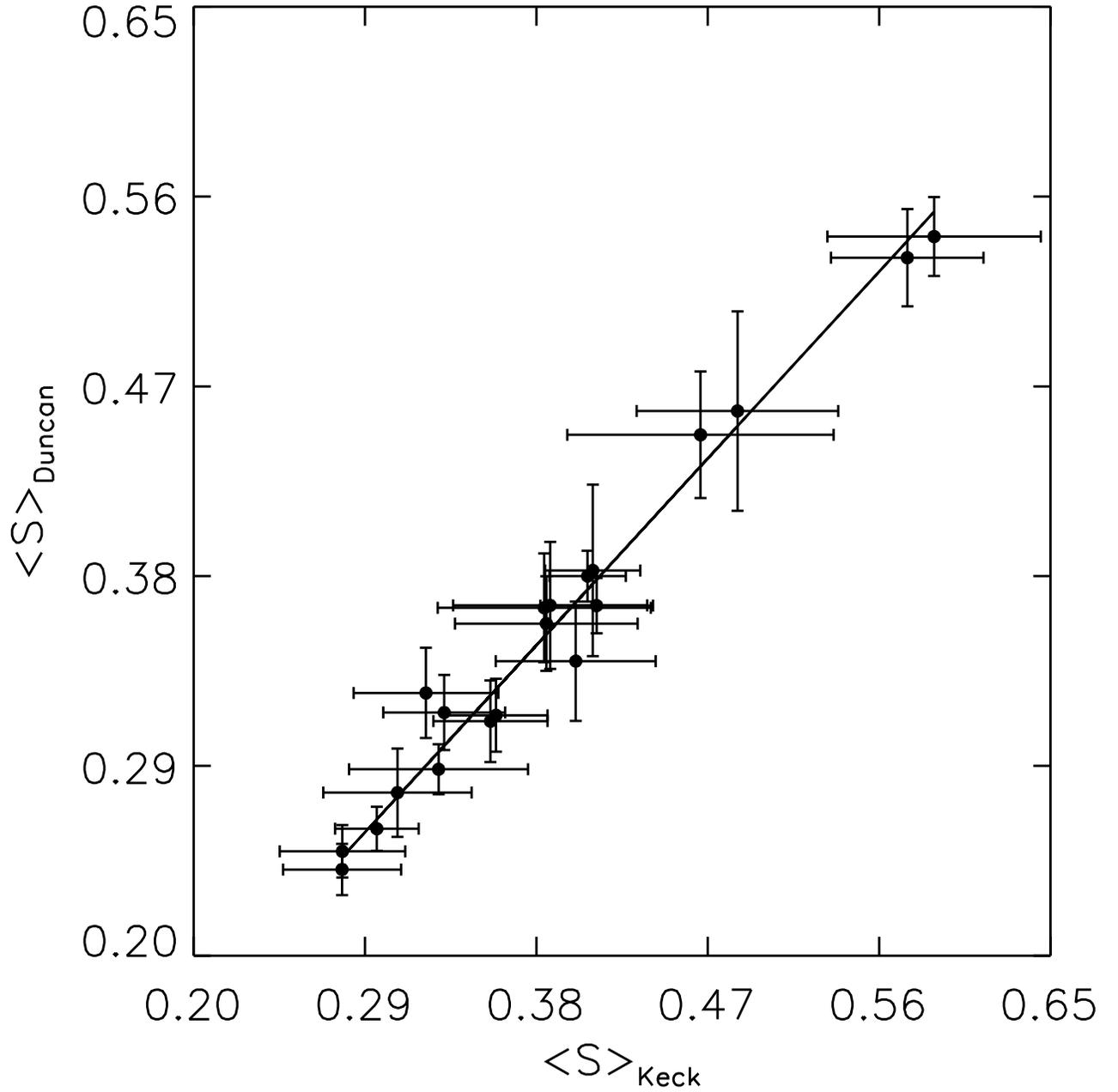}
\caption{Comparison of our computed S index with published values from
\citet{DuFrLa84}. The straight line is the linear fit discussed
in section 2.3. Error bars indicate the rms about the mean S value for each
star.}
\end{figure}

\begin{figure}
\plotone{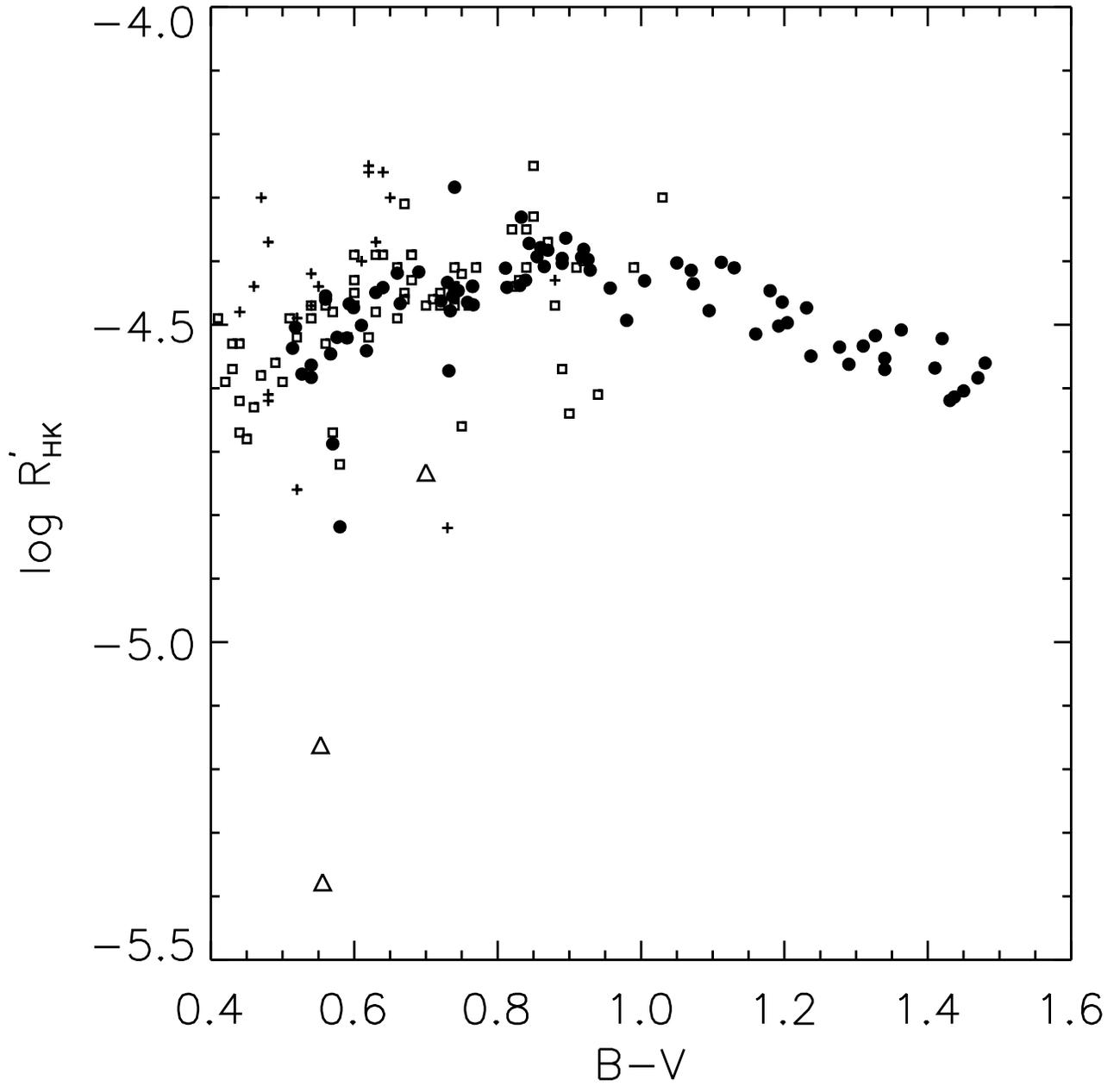}
\caption{R$^{\prime}_{HK}$ vs. B-V. Filled circles=Our Hyades program stars. 
Open triangles= Members according to \citet{PeBrLe98} but are non-members due 
to high Hipparcos distances, photometry below 
the main-sequence and velocities which disagree with members according to 
D. Latham (private communication). Open 
squares=\citet{Ru87} Hyades data. Crosses=\citet{Ru87} Pleiades data.}
\end{figure}

\begin{figure}
\plotone{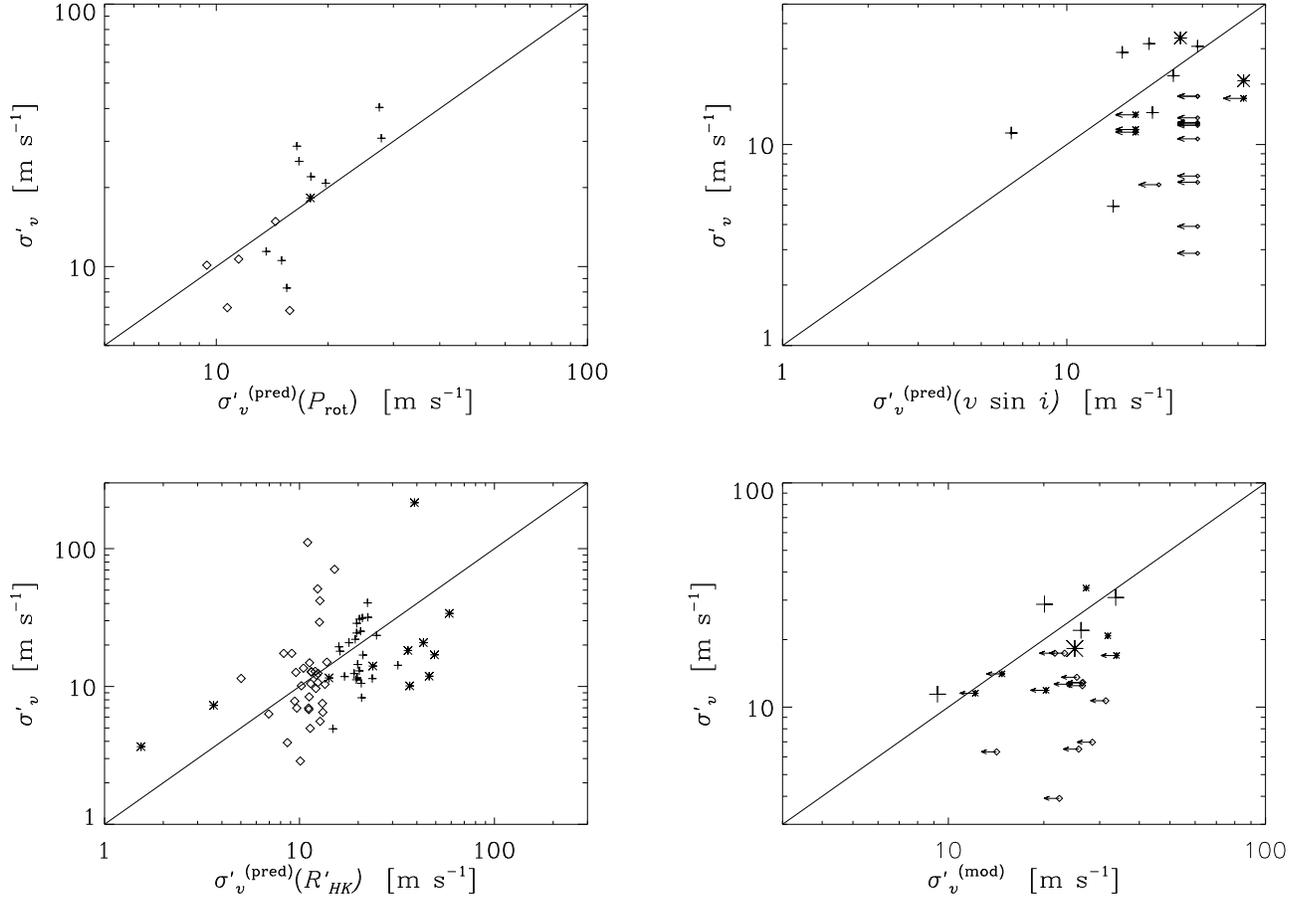}
\caption{Comparison between measured $\sigma'_v$ for our stars, and
predicted $\sigma'_v$  based on empirical based on relations (see SBM) between
$\sigma'_v$ and $P_{\rm rot}$ (top left), $v \sin i$ (top right) and
$R'_{\rm HK}$ (bottom left; separate fits for G and K stars used).
Bottom right shows comparison between $\sigma'_v$(Hyades) and
those predicted by the simple spot/plage model in SBM (smaller symbols
have estimated $P_{\rm rot}$).
In each case, F, G, and K stars are given by $*$, $+$, and diamonds,
respectively, and the solid line indicates the observed $\sigma'_v$ =
$\sigma^{\prime \rm (pred.)}_v$.
}
\end{figure}

\begin{figure}
\plotone{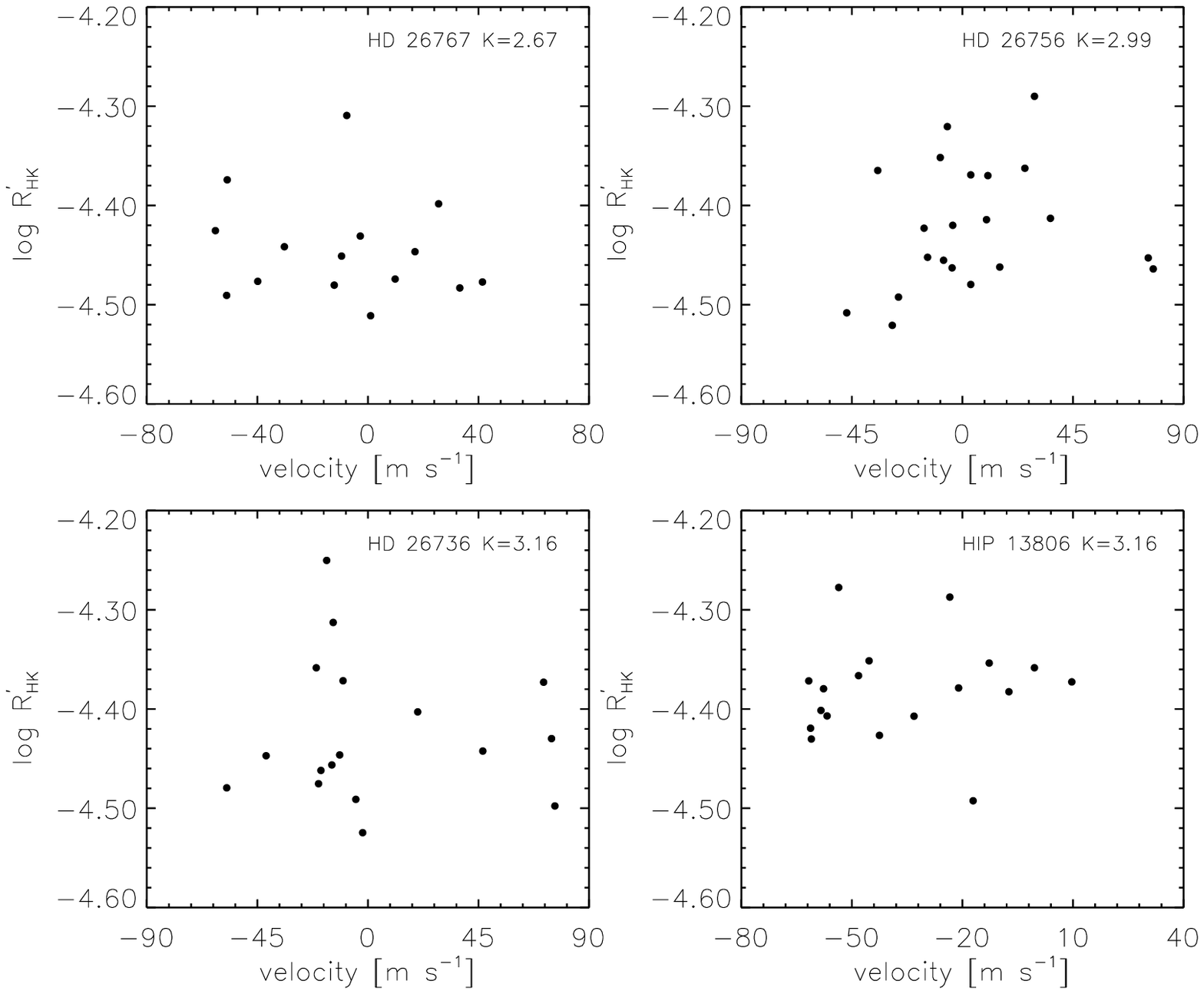}
\caption{Examples of stars showing uncorrelated trends in
logR$^{\prime}_{HK}$ vs. radial velocity.}
\end{figure}

\begin{figure}
\plotone{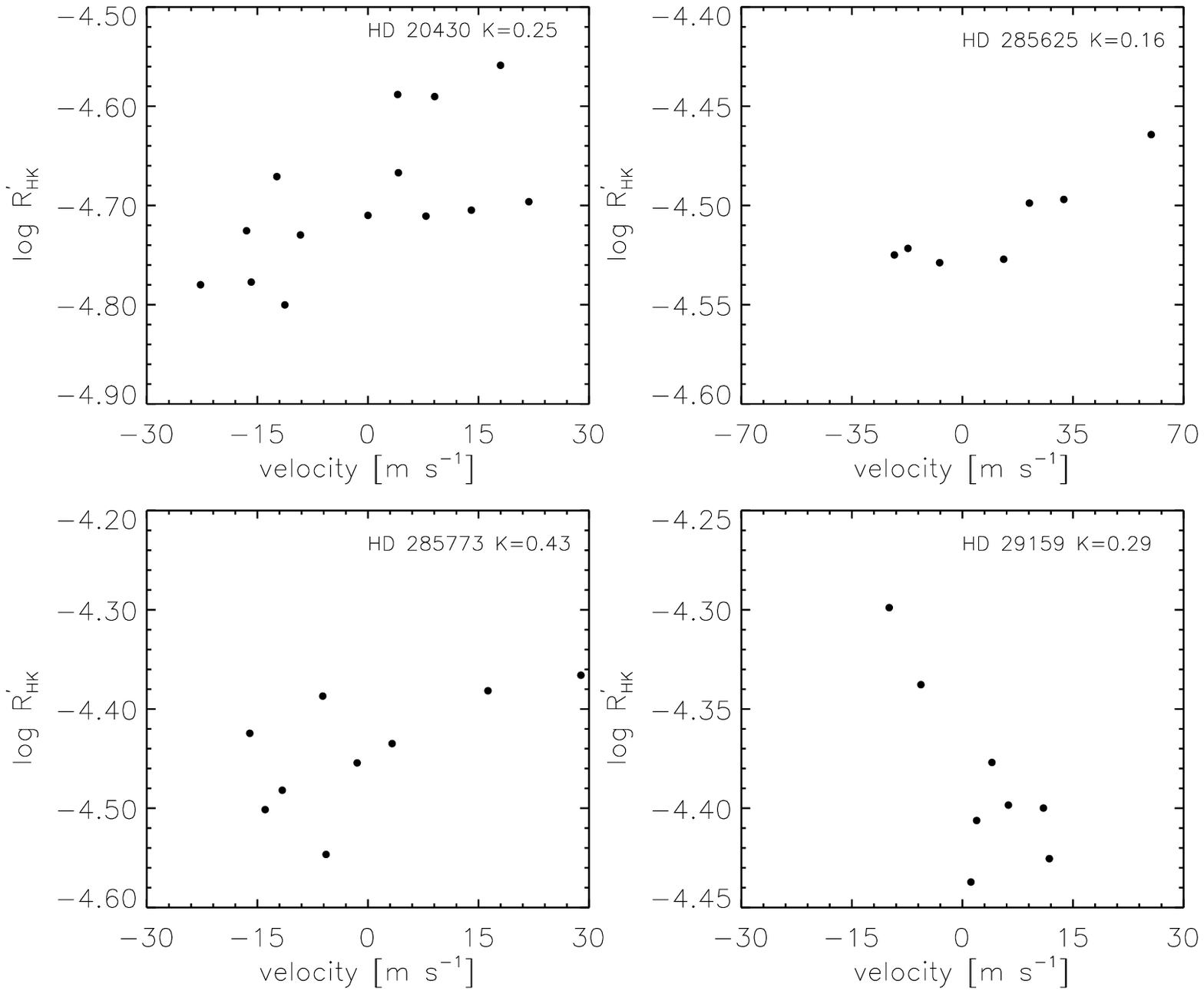}
\caption{Examples of stars showing correlated trends in
logR$^{\prime}_{HK}$ vs. radial velocity.}
\end{figure}

\clearpage
\begin{deluxetable}{ccccr}
\tabletypesize{\scriptsize}
\tablecaption{Internal Errors versus Rotation for the Program Stars\label{tbl-1}}
\tablewidth{0pt}
\tablehead{
\colhead{HD} & \colhead{Other Name} & 
\colhead{$v \sin i$} &
\colhead{$\langle \sigma_i \rangle$} &\colhead{Reference} \\
\colhead{ } & & [km s$^{-1}$] & [m s$^{-1}$&] 
}
\startdata
18632 & BD+07 459 & 1.8$\pm$1.0 & 3.52$\pm$0.72 & 1 \\
26756 & vB 17 & 6.4$\pm$3.0 & 4.94$\pm$0.69 & 2\\
26736 & vB 15 & 7.0$\pm$0.7 & 5.13$\pm$0.99 & 3\\
27282 & vB 27 & 6.6$\pm$3.0 & 3.81$\pm$0.45 & 2 \\
27406 & vB 31 & 12.0$\pm$1.0 & 6.19$\pm$1.22 & 3\\
27859 & vB 52 & 8.0$\pm$ 0.7 & 5.76$\pm$0.80 & 3\\
20899 & vB 64 & 5.0$\pm$0.7 & 3.81$\pm$0.46 & 3\\
28205 & vB 65 & 9.0$\pm$3.0 & 7.02$\pm$0.77 & 4\\
28237 & vB 66 & 8.0$\pm$3.0 & 6.80$\pm$0.90 & 4\\
28344 & vB 73 & 7.0$\pm$0.7 & 6.80$\pm$1.39 & 3\\

\enddata
\tablerefs{
(1) \citet{Fe97};We adopted an error of 1.0 km s$^{-1}$. 
(2) \citet{StGrSc00}; We adopted errors of 3.0 km s$^{-1}$.
(3) \citet{So82}
(4) \citet{Kr65}; We adopted errors of 3.0 km s$^{-1}$. }
\end{deluxetable}

\begin{deluxetable}{lccccc}
\tabletypesize{\scriptsize}
\tablecaption{Comparison of Observed and Predicted/Modeled $\sigma'_v$
\label{tbl-2}}
\tablewidth{0pt}
\tablehead{
\colhead{Type of} & \colhead{Spec. type} & $N$ &
\colhead{$\langle \sigma^{\prime \rm (pred.)}_v-\sigma'_v\rangle$}   &
\colhead{$\sigma_{rms}$}& \colhead{$\sigma_{rms}$(Lick)\tablenotemark{2}}\\
\colhead{Fit/Model\tablenotemark{1}} & \colhead{studied} &  &
[dex] &  [dex] & [dex]
}
\startdata

$R^{\prime}_{\rm HK}$(F) & F & 11 & 0.124 &  0.444 & 0.41 \\
$R^{\prime}_{\rm HK}$(F) & F\tablenotemark{3} & 10 & 0.210 &  0.376 & 0.41 \\
$R^{\prime}_{\rm HK}$(G) & F\tablenotemark{3} & 10 & -0.107 & 0.213 & ... \\
$R^{\prime}_{\rm HK}$(G+K) & F\tablenotemark{3} & 10 & -0.213 & 0.284 & ... \\
$R^{\prime}_{\rm HK}$(G+K) & G & 24 & -0.045  & 0.216 & ... \\ 
$R^{\prime}_{\rm HK}$(G+K) & K & 34 & 0.117 &  0.348 & ... \\ 
$R^{\prime}_{\rm HK}$(G+K) &K\tablenotemark{3} &33 & 0.146 & 0.320 & ... \\ 
$R^{\prime}_{\rm HK}$(G+K) & G+K & 58 & 0.050  & 0.298 & 0.25 \\ 
$R^{\prime}_{\rm HK}$(G+K) & G+K\tablenotemark{3} & 57 & 0.065 & 0.278 & 0.25 \\
$R^{\prime}_{\rm HK}$(G) & G & 24 & 0.073 &  0.224 & ... \\ 
$R^{\prime}_{\rm HK}$(K) & K & 34 & -0.040 &  0.329 & ... \\ 
$R^{\prime}_{\rm HK}$(K) &K\tablenotemark{3} & 33 & -0.011 & 0.284 & ... \\ 
$v \sin i$(F) & F & 3 & 0.125  &  0.263  & 0.31 \\
$v \sin i$(G+K) & G\tablenotemark{4} & 7 & -0.016  & 0.265  & 0.22 \\ % GK fit
$P_{\rm rot}$(F) & F & 1 & -0.008  & ... & 0.35 \\
%$P_{\rm rot}$(G) & G & 9 & -0.027  & 0.170 & ...$^*$ \\
%$P_{\rm rot}$(K) & K & 5 & 0.108  & 0.206 & ...$^*$ \\ 
$P_{\rm rot}$(G+K) & G+K & 14 & 0.021 & 0.176 & 0.22 \\
model & F+G\tablenotemark{4} & 5 & 0.001  & 0.121 & 0.14 \\
model\tablenotemark{5} & F+G\tablenotemark{4} & 7 & 0.013  & 0.130 & 0.16 \\
\enddata
\tablenotetext{1}{e.g., $R^{\prime}_{\rm HK}$(G+K) indicates that the
$\sigma^{\prime \rm (pred.)}$ based on the
empirical fit for G and K stars of $\sigma'_v$(Lick) vs. $R^{\prime}_{\rm HK}$
(from SBM) is compared
to Hyades stars of the spectral type listed in column (2). Note:
separate G star ($\propto {R^{\prime}_{\rm HK}}^{1.15}$) and K star
($\propto {R^{\prime}_{\rm HK}}^{1.19}$) fits for $R^{\prime}_{\rm HK}$
based on Lick data were not listed in SBM.}
\tablenotetext{2}{from SBM}
%, or if marked with $*$, calculated using data from SBM}
\tablenotetext{3}{with one outlier removed}
\tablenotetext{4}{No Hyad K dwarfs in our sample have published $v \sin i$ 
measurements}
\tablenotetext{5}{Assuming $f_S = 0$ for F stars}
\end{deluxetable}

{\setlength{\textheight}{25cm}
\begin{deluxetable}{ccrrrrrrr}
\tabletypesize{\scriptsize}
\tablecaption{Program Information \label{tbl-3}}
\tablewidth{0pt}
\tablehead{
\colhead{HD or HIP $\#$} & \colhead{Other Name} & \colhead{$\#$ obs.}   &
\colhead{B-V  } & \colhead{$\langle R^{\prime}_{\rm HK} \rangle$} &
\colhead{$\sigma_{<R^{\prime}_{\rm HK}>}$}  & 
\colhead{$r$} & 
\colhead{$P_{c}(R^{\prime}_{\rm HK},r)$} &
\colhead{$K(r^{2})$}   
}
\startdata
HD 14127  & BD+04 378& 12 & 0.567 & 2.84 & 0.38 & 0.55 & 0.07  & 0.66 \\ 
HIP 13600 & BD+17 455 & 10 & 0.704 &  1.85 & 0.31 &-0.08&0.83  & 2.41 \\
HIP 13806 &vB 153  & 18 & 0.855&  4.18 & 0.49 &  0.09 & 0.73  & 3.16 \\ 
HD 18632  &BD+07 459& 9 &  0.926&  4.00 & 0.52 & 0.00 & 1.00  & 2.33 \\
HD 19902  &BD+32 582& 10 &0.732 &   2.67&  0.40 & 0.07& 0.85  & 2.43 \\
HD 20430  &vB 1    & 13 & 0.567 &  2.07&  0.37 & 0.62 & 0.02  & 0.25 \\
HD 20439  &vB 2    & 13 & 0.617 &  2.88&  0.31 & 0.30 & 0.33  & 1.79 \\
HIP 15563 &BD+07 499& 9 & 1.130 &  3.89&  0.38 & -0.60& 0.09  & 0.62 \\
HIP 15720 &\nodata & 8 &  1.431 & 2.40 & 0.18 & 0.10  & 0.82  & 2.14 \\ 
HIP 16529 &vB 4    & 9 &  0.844 &  4.25 & 0.36 & 0.13 & 0.73  & 2.21 \\
HIP 16908 &vB 5    & 10 &  0.917 &  4.04 & 0.38 &0.32 & 0.36  & 1.77 \\
HD 23453 &BD+25 613 & 8  & 1.437&  2.43 & 0.22 & -0.27& 0.51  & 1.87 \\
HIP 17766 &G7-15   & 7 &  1.340 &  2.69 & 0.29 &  0.02& 0.96  & 2.03 \\
HIP 18018 & vB 170  & 11 & 1.160 &   3.06 & 0.36 & 0.30& 0.38  & 1.79 \\
HD 286363 &BD+12 524& 7 & 1.070 &   3.85 & 0.45 & 0.06& 0.90  & 2.02 \\
HD 25825 &vB 7    & 8 &   0.895 & 4.33 & 0.36 & 0.19  & 0.65  & 2.03 \\
HIP 18946 &BD+19 650 & 7 & 1.095 & 3.33 & 0.47 &0.23 & 0.62  & 1.82 \\
HIP 19082 &L 12    & 7 &  1.347 &  2.80 & 0.29  & 0.61& 0.15  & 0.81 \\
HD 285367 &BD+17 679 & 8 & 0.890&    4.02 & 0.24 & 0.29& 0.48 & 1.82 \\
HD 25825 &vB 10   & 8 &  0.593 &  3.41 & 0.47 & 0.20 &  0.63  & 2.01 \\
HD 285507 &J 231   & 8 &  1.180 & 3.58&  0.35 & 0.41 &  0.31  & 1.50 \\
HIP 19261B &vB 12   & 8 & 0.740 &  5.20 & 0.59 & -0.01 & 0.98 & 2.18 \\
 HD 285482 &BD+16 558 & 7 &1.005 &   3.71 & 0.29 & -0.30& 0.52 & 1.69 \\
HD 286554 &J 233   & 7 &  1.327 &  3.04 & 0.24 & 0.11  & 0.82 & 1.99 \\
HD 26257 &BD-00 648 & 9 & 0.553 &  0.69&  0.20 & -0.51 & 0.16 & 0.95 \\
HIP 19441 &BD+08 642 & 6& 1.192 &   3.15&  0.29 & 0.33 & 0.52  & 1.67\\
HD 26756 &vB 17   & 21 & 0.693  & 3.83&  0.57 & 0.14  & 0.53  &2.99 \\
HD 26767 &vB 18   & 15 & 0.640  & 3.62&  0.47 &  -0.15 & 0.60 & 2.67\\
HD 26736 &vB 15   & 17 & 0.657 &  3.81&  0.68 & -0.06  & 0.82  & 3.16 \\
HD 26784 &vB 19   & 11 & 0.514  & 2.90&  0.42 &  -0.12& 0.73  & 2.45 \\
HD 286589 &vA 68   & 7 & 1.204  & 3.18&  0.44 & 0.09  & 0.85  & 2.00 \\
HD 285625 &vA 72   & 8 & 1.363  & 3.06&  0.21 & 0.86  & 0.01  & 0.16 \\
HD 285590 &vA 75   & 7 & 1.290 &  2.74&  0.16 & 0.17  & 0.71  & 1.92 \\
\nodata   &vA 115  & 6 & 1.470 &  2.61&  0.31 & 0.73 & 0.10   & 0.87 \\
HD 285690 &vB 25   & 7 & 0.980 &  3.21&  0.23 & 0.39 & 0.39   & 1.48 \\
\nodata  &vA 146  & 6  & 1.420 & 3.01 & 0.34  & 0.56 & 0.24   & 1.27 \\
HD 27250 &vB 26   & 8  & 0.745 & 3.58 & 0.35 & -0.07 & 0.87   & 2.16\\
HD 27282 & vB 27   & 9  & 0.721 & 3.45 & 0.64 & 0.53  & 0.14   & 0.86 \\
HD 27406 &vB 31   & 8  & 0.560 & 3.51 & 0.79 &0.31  & 0.46  &1.79 \\ 
HD 27732 &vB 42   & 8  & 0.758 & 3.43 & 0.23 & 0.33  & 0.42  & 1.73 \\
HIP 20485 &vB 173  & 6 & 1.231 &  3.36&  0.20 &-0.07 & 0.89   &  1.86 \\
HD 27771 &vB 46   & 8  & 0.855 & 4.05 & 0.33  & -0.24& 0.56   &  1.94 \\
HD 27835 &vB 49   & 8 &  0.590 & 3.01 & 0.62 &  -0.20& 0.63  &  2.01\\
HD 27808 &vB 48   & 8 &  0.518 & 3.13 & 0.38 &  0.28 &  0.50  & 1.85 \\
HIP 20485 &vB 174  & 7 & 1.050 &  3.96 & 0.50 &  0.08 & 0.86   & 2.01 \\
HD 27859 &vB 52   & 9 &  0.599 & 3.36 & 0.50 &  0.14 &  0.73  & 2.20 \\
\nodata &vA 354  & 7 &  1.310 & 2.93 & 0.40 & 0.41  & 0.35    &1.39 \\
\nodata &vA 383  & 6 &  1.450 & 2.49 & 0.20  &0.03  & 0.95    &1.87 \\
HD 20899 &vB 64   & 9 & 0.664  & 3.41 & 0.39 & 0.14  & 0.72   & 2.19 \\
HD 28205 &vB 65   & 8 & 0.537  & 2.61 & 0.44 & 0.10  & 0.82  & 2.15 \\
HD 28237 &vB 66   & 8 & 0.560  & 3.47 & 0.63 & 0.26  & 0.53  & 1.89 \\
HD 285830 &vB 179  & 8 & 0.929 &  3.85&  0.32& 0.35  & 0.40  & 1.69 \\
HD 28258 &vB 178  & 9 &  0.839 & 3.72 & 0.36 & -0.48 & 0.19  &  1.07 \\
HD 28344 &vB 73   & 10 & 0.609 &  3.15&  0.60 & 0.21  & 0.56  & 2.15 \\
\nodata &vA 502  & 6 &  1.410 & 2.70 & 0.15  & -0.47 & 0.35  &  1.47 \\
HD 283704 &vB 76   & 10 & 0.766 &  3.40&  0.40&-0.12 & 0.73  &  2.35 \\
HD 285773 &vB 79   & 9 & 0.831 &  3.64&  0.49 &0.66  & 0.06  & 0.43 \\
HD 28462 &vB 180  & 7  & 0.865 & 3.90 & 0.63 &  -0.39& 0.39  &   1.46 \\
HD 28593 &vB 87   & 9  & 0.734 & 3.32 & 0.68 &0.11  & 0.77   &2.24 \\
HD 28635 &vB 88   & 10 & 0.540 &  2.73&  0.57 & 0.69 & 0.03  &  0.34 \\
\nodata &vA 637  & 6  & 1.480 & 2.75  &0.18 & -0.52  & 0.29  & 1.36 \\
HD 28805 &vB 92   & 7  & 0.740 & 3.59 & 0.29 &  0.09 & 0.84  &  2.00 \\
HD 284552 &L 66    & 6 & 1.237  & 2.82&  0.41 & 0.55 & 0.26  &  1.31 \\
HD 28878 &vB 93   & 8  & 0.890 & 3.95 & 0.56  & -0.39& 0.34  &   1.56 \\
HD 285837 &L 65    & 6 & 1.197  & 3.43&  0.51 & 0.77 & 0.07  &  0.77 \\
HD 28977 &vB 183  & 7  & 0.920 & 4.16 & 0.61  & 0.06 & 0.90  &  2.02 \\
HD 28892 &vB 97   & 8  & 0.631 & 3.55 & 0.49  & -0.29& 0.49  &   1.84 \\
HD 29159 &vB 99   & 8  & 0.870 & 4.14 & 0.46  & -0.80& 0.02  &  0.29 \\
HD 29419 &vB 105  & 8  & 0.576 & 3.02 & 0.78 & -0.24 & 0.56  &  1.93 \\
HD 286929 &J 311   & 6 & 1.073  & 3.67 & 0.35 & 0.71 & 0.12  &  0.93 \\
HD 284574 &vB 109  & 7 & 0.811 &  3.88&  0.71 & -0.21& 0.65  &   1.86 \\
HIP 22177 &J 326 & 6 &  1.277 & 2.91&  0.09 &0.70  & 0.12  &  0.95 \\
HD 284653 &J 330   & 6 & 1.112 &  3.96&  0.23 &-0.45& 0.38   &  1.50 \\
HD 30505 &vB 116  & 9  & 0.833 & 4.67 & 0.88 &  -0.39& 0.30  &   1.42 \\
HD 30589 &vB 118  & 10 & 0.578  & 1.52&  0.28 &-0.24& 0.50   &  2.04 \\
HD 30809 &vB 143  & 7  & 0.527 & 2.64 & 0.50 &-0.29 & 0.54   & 1.72\\ 
HD 31609 &vB 127  & 7  & 0.737 & 3.51 & 0.50  &-0.10& 0.84   &  1.99 \\
\nodata &BD+04 810 & 6 & 0.957 &  3.61&  0.23 &-0.43& 0.39   &  1.52 \\
HD 32347 &vB 187  & 7  & 0.765 & 3.64 & 0.57  &-0.12& 0.80   &  1.98 \\
HD 240648 & BD+17 841& 7 &0.730 &   3.69&  0.24& 0.25& 0.59   &  1.79 \\
HD 242780 &BD+11 772& 6 & 0.765 &  3.63&  0.23 &0.81& 0.05   &  0.63 \\
HD 35768 &BD+32 955& 6 & 0.556 &  0.42 & 0.17 &0.16 & 0.76   & 1.82 \\
 \enddata
\end{deluxetable}
}
%end{landscape}
\end{document}